\documentclass[amsmath,amssymb,aps,prl,twocolumn,showpacs,reprint,superscriptaddress,longbibliography]{revtex4-1}

\newcommand{\etal}{\textit{et al}.}
\usepackage{footmisc}
\usepackage{todonotes}
\usepackage{textcomp}
\usepackage{graphicx}
\usepackage{amssymb,amsfonts,amsmath}
\usepackage{bm}
\usepackage{textcomp}
\usepackage[english]{babel}
\usepackage{url}
\usepackage{hyperref}
\usepackage{url} 

\begin{document}

\title{Scalable, symmetric atom interferometer for infrasound gravitational wave detection}

\def\affiqo  {\affiliation{Leibniz Universit\"at Hannover, Institut f\"ur Quantenoptik, Welfengarten 1, 30167 Hannover, Germany }}
\def\affulm {\affiliation{Institut f\"ur Quantenphysik and Center for Integrated Quantum Science and Technology \textnormal{(IQ$^{\text{ST}}$)}, Universit\"at Ulm, Albert-Einstein-Allee 11, D-89081 Ulm, Germany}}

\author{C. Schubert}
\email[]{schubert@iqo.uni-hannover.de}
\author{D. Schlippert}
\author{S. Abend}
\affiqo
\author{E. Giese}
\affiliation{Department of Physics, University of Ottawa, 25 Templeton Street, Ottawa, Ontario, Canada K1N 6N5}
\affulm
\author{A.~Roura}
\affulm
\author{W.~P.~Schleich}
\affulm
\affiliation{Institute of Quantum Technologies, German Aerospace Center (DLR), D-89069 Ulm, Germany}
\affiliation{Institute for Quantum Science and Engineering \textnormal{(IQSE)}, Texas A\&M AgriLife Research and Hagler Institute for Advanced Study, Texas A\&M
University, College Station, Texas 77843-4242, USA}
\author{W. Ertmer}
\author{E. M. Rasel}
\affiqo

\date{\today}

\begin{abstract}
We propose a terrestrial detector for gravitational waves with frequencies between 0.3\,Hz and 5\,Hz.
Therefore, we discuss a symmetric matter-wave interferometer with a single loop and a folded triple-loop geometry. 
The latter eliminates the need for atomic ensembles at femtokelvin energies imposed by the Sagnac effect in other atom interferometric detectors.
It also combines several advantages of current vertical and horizontal matter wave antennas and enhances the scalability in order to achieve a peak strain sensitivity of $2\cdot10^{-21}\,/\sqrt{\mathrm{Hz}}$.
\end{abstract}

\maketitle
The direct observation of gravitational waves with laser interferometers~\cite{Abbott2016PRL_1, Abbott2016PRL_2} marks the beginning of a new area in astronomy with new searches targeting signals in a broader frequency band with a variety of detectors.
One class of proposed detectors relies on atom interferometers rather than macroscopic mirrors as inertial references.
Indeed, vertical~\cite{Dimopoulos2008PRD,Coleman2018arXiv,magiswebsite} or horizontal~\cite{Canuel2017arXiv} baselines interrogated by common laser beams propagating along those baselines have recently been proposed for terrestrial observatories.
In this letter, we present a new class of symmetric atom interferometers enabling single and multi-loop geometries for broad and narrow-band detection in the frequency range of 0.3\,Hz to 5\,Hz.

Today's detectors are based on laser interferometers and operate in the acoustic frequency band between ten and hundreds of hertz~\cite{Abbott2016PRL_1, Abbott2016PRL_2, Acernese2015CQG, Grote2010CQG}.
Future space-borne interferometers, such as LISA~\cite{Robson2018arXiv,AmaroSeoane2017arXiv,Danzman2013arXiv,Amaro-Seoane2012CQG}, are designed to target signals in the range of millihertz to decihertz.
Detectors operating in the mid-frequency band~\cite{Kawamura2011CQG} or improving on the frequency band of current ground based devices~\cite{Punturo2010CQG} were also proposed and investigated.

Matter-wave interferometers~\cite{Canuel2017arXiv,Hogan2016PRA,Graham2013PRL,Yu2011GRG,Hogan2011GRG,Hohensee2011GRG,Dimopoulos2008PRD}, mechanical resonators~\cite{Aguiar2011RAA,Weber1960PR,Weber1969PRL}, and optical clocks~\cite{Kolkowitz2016PRD} are pursued to search for sources of gravitational waves in the infrasound domain featuring frequencies too low to be detected by today's ground-based detectors.
Waves in this band are emitted for example by inspiraling binaries days or hours before they merge within fractions of a second~\cite{Sesana2017JPCS,Cutler2016arXiv} as the first observed event GW150914~\cite{Abbott2016PRL_1,Abbott2016PRL_2}.
Hence, terrestrial observation in this band could be combined with standard astronomical observations.
The vast benefits of joint observations manifested itself in the case of a merger of neutron stars~\cite{Abbott2017AJL}.

Our detector is based on a novel type of interferometer where matter waves form a single or several folded loops of symmetric shape~\footnote{Kindred concepts~\cite{Dubetsky2006PRA} were discussed for space-borne detectors~\cite{Hogan2011GRG}.}.
An antenna employing folded multi-loop interferometers shows three distinct advantages: (i) The detector is less susceptible to environmental perturbations, and less restrictive on the expansion rate of the atomic ensemble, which otherwise needs to be at the energy level of femtokelvins. (ii) It combines the scalability of arm lengths of horizontal and the single-laser link of vertical antenna types, (iii) and it is less susceptible to technical noise such as the pointing jitter of the atomic sources.

\begin{figure}[tp]
\centering
\includegraphics[width=\columnwidth]{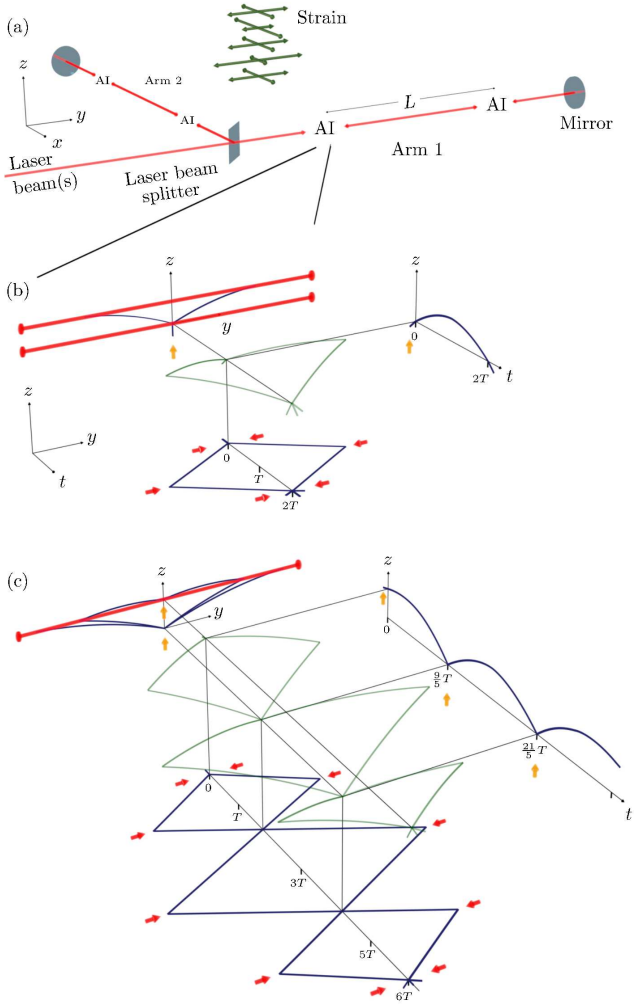}
\caption{(Color online) (a) Matter-wave detector for infrasound gravitational waves. 
Two horizontal arms (along $x$ and $y$) each with at least two atom interferometers are sensitive to the phase of the light beams (red lines).
It is maximally sensitive to gravitational waves travelling along $z$, and periodically stretching and squeezing the detector arms.
(b) Spacetime diagram and geometric projections of a symmetric single-loop atom interferometer with a pulsed twin-lattice formed by retroreflection. Atoms are launched vertically along $z$ (orange arrows in the $z(y)$ and $z(t)$ diagrams) by a coherence preserving mechanism such as Bloch oscillations combined with double Bragg diffraction~\cite{Abend2016PRL}.
The first interaction with the bottom twin-lattice at $t=0$ excites the atoms into a coherent superposition of two opposite momenta as shown in the $z(y)$ and $y(t)$ diagrams.
The second interaction at the apex at $t=T$ inverts the momenta and, finally, a recombination pulse at the bottom lattice at $t=2T$ enables interference.
(c) After exciting the atoms into a coherent superposition of two momenta at $t=0$, three subsequent interactions each inverting the momenta at $t=T$, $t=3T$, $t=5T$, form three loops before the recombination at $t=6T$.
The two vertical relaunches at $t=(9/5)T$ and $t=(21/5)T$ fold the interferometer geometry which requires only a \textit{single} horizontal beam splitting axis for manipulating the atoms as opposed to (b).
}
\label{fig:GWD_3_pics}
\end{figure}

In this Letter, we explain the concept of our detector, and compare a symmetric single-loop interferometer with a folded multi-loop one, confronting their scaling, spectral responses, and critical parameters.
Similar to the matter wave interferometric gravitational antenna MIGA~\cite{Canuel2017arXiv,Chaibi2016PRD} and laser interferometers~\cite{Abbott2016PRL_1, Abbott2016PRL_2} our detector concepts have two perpendicular, horizontal arms, depicted in Fig.~\ref{fig:GWD_3_pics}~(a), suppressing laser frequency noise at lower frequencies as in light interferometers~\cite{AmaroSeoane2017arXiv,Abbott2016PRL_1,Acernese2015CQG}.
Two light-pulse atom interferometers, separated by a distance $L$, and located in each arm, are sensitive to the phase of the light pulses travelling along the $x$ and $y$ axis between the interferometers and are employed for coherent manipulation of the atoms.
A gravitational wave, propagating along the $z$ axis, modifies the geodesic of the light connecting the interferometers and modulates its phase which appears in the differential signal of the two atom interferometers.

In contrast to Refs.~\cite{Chaibi2016PRD,Dimopoulos2008PRD}, our atom interferometers are symmetric~\cite{Gebbe2019arXiv,Ahlers2016PRL,Hogan2011GRG,Leveque2009PRL} as an identical number of photon recoils is transferred to both paths of the atom interferometer during beam splitting, deflection, and recombination.
For this purpose we employ a twin-lattice, i.e. two counter-propagating lattices formed by retroreflecting a light beam with two frequency components.
Neglecting the non-vanishing pulse duration of the atom-light interaction~\cite{Bonnin15PRA,Cheinet2008IEEE,Antoine07PRA}, the phase difference~\cite{Hohensee2011GRG,Dimopoulos2008PRD} in the single-loop atom interferometer of Fig.~\ref{fig:GWD_3_pics}~(b) induced by a gravitational wave of amplitude $h$ and angular frequency $\omega{\equiv}2\pi f$ reads
\begin{equation*}
    \phi_{\mathrm{SL}}=2\,h\,k\,L\left[\mathrm{cos}(2\pi f\,T)-1\right].
\end{equation*}

Hence, three parameters determine the spectral sensitivity of such a detector: (i) The distance  $L$ between the interferometers, (ii) the number of photon momenta transferred differentially between the two interferometer paths corresponding to the wave number $k$, and
(iii) the time $T$ during which the atoms are in free fall.
The ability to change $T$ in this setup allow for adjusting the frequency of the maximum spectral response which scales with the inverse of $T$.

Unfortunately, spurious effects can mask the phase induced by a gravitational wave.
For example, in the single-loop interferometer the shot-to-shot jitter of the initial velocity $\mathbf{v}$ and position $\mathbf{r}$ causes noise as it enters the phase terms $2\left(\mathbf{k}\times\mathbf{v}\right)\mathbf{\Omega}T^2$, $\mathbf{k}\,\Gamma\,\mathbf{r}\,T^2$, and $\mathbf{k}\,\Gamma\,\mathbf{v}\,T^3$ ~\cite{Hogan2008arXiv,Borde2004GRG} arising from the Sagnac effect and gravity gradients.
Here $\mathbf{\Omega}$ and $\Gamma$ denote the Earth's angular velocity and the gravity gradient tensor~\footnote{The nonzero duration of the atom-light interaction is neglected in the derivation of the phase terms since it does not qualitatively change the result.}.
Whereas the impact of the gravity gradients can be compensated~\cite{Roura2014NJP,Roura2017PRL}, typical parameters targeting a competitive strain sensitivity enforce kinetic expansion energies of the atomic ensembles to suppress the spurious Sagnac effect corresponding to femtokelvin temperatures.
Such a regime is beyond reach even with delta-kick collimation~\cite{Rudolph2016Diss,Kovachy2015PRL,Muntinga2013PRL}.

This challenge can be mitigated by a suitably tailored geometry with atomic trajectories following multiple loops inside the interferometer~\cite{Hogan2011GRG,Marzlin1996PRA} as depicted in Fig.~\ref{fig:GWD_3_pics}~(c).
In this scheme the interferometer begins after a vertical launch at time $t=0$, with the first interaction with the symmetric twin lattice creating a coherent superposition of two momentum states.
On their way downward, the atoms are coherently reflected at $t=T$ such that they complete their first loop and cross each other at the bottom at $t=(9/5)T$.
Here, they are launched upwards again~\cite{Abend2016PRL} and their horizontal motion is inverted at the vertex in the middle of the interferometer at $t=3T$.
After completion of the second loop, they are again relaunched at $t=(21/5)T$, once more redirected at $t=5T$, and finally recombined at $t=6T$ to close the interferometer.
In order to suppress spurious phase contributions due to the Sagnac effect and gravity gradients, the wave numbers of the five pulses are adjusted~\cite{Hogan2011GRG} to $k$, $\frac{9}{4}k$, $\frac{5}{2}k$, $\frac{9}{4}k$, $k$, corresponding to the differential momentum between the two paths of the interferometer.

Compared the symmetric single-loop interferometer, the adjusted wave numbers and additional loops change the phase difference between the two interferometers~\cite{Hogan2011GRG} which now reads 
\begin{equation*}
  \phi_{\mathrm{FTL}}=\frac{1}{2}h\,k\,L\left[5-9\cos\left(2\cdot2\pi f\,T\right)+4\cos\left(3\cdot2\pi f\,T\right)\right]
\end{equation*}
for our folded triple-loop sequence.
This scheme shares the advantage of a horizontal detector~\cite{Canuel2017arXiv,Chaibi2016PRD} enabling us to operate the interferometers separated by up to several ten kilometers similar to light interferometers~\cite{Abbott2016PRL_1,Abbott2016PRL_2}.
Moreover, it requires only a \textit{single} horizontal beam axis in each arm for the coherent manipulation of the atoms.
Hence, $T$ is not constrained by the difference in height between the upper and lower twin lattice, and can be tuned to specific frequencies of the gravitational waves similar to vertical antennas, where the light beam propagates along the direction of the free fall of the atoms.

Finally, the folded triple-loop interferometer mitigates several important drawbacks of other concepts.
For example, the specific combination of three loops and increased momentum transfer by the central pulses compared to the initial and final pulse, renders this interferometer insensitive to fluctuations of initial position and velocity~\cite{Hogan2011GRG}, mitigating the requirement for femtokelvin energies.

Unfortunately, folded multi-loop interferometers are susceptible to pointing noise of the relaunch vector.
In a model with a vanishing atom-light interaction time and no timing errors, the requirements remain comparable to those for a single-loop geometry, and have to be limited to picoradians.
For relaunches not centered on the intersections of the trajectories (see Fig.~\ref{fig:GWD_3_pics}~(c)), new terms appear if their directions are not properly aligned~\footnote{See Supplemental Material at [URL will be inserted by publisher] for the impact of timing errors.}.
Table~\ref{tab:parameter} summarizes the requirements imposed by the two geometries.
The impact of mean field effects is discussed in Ref.~\footnote{See Supplemental Material at [URL will be inserted by publisher] for the impact of relative density fluctuations between the two interferometer arms}.

\begin{figure}[tp]
\centering
\includegraphics[width=\columnwidth]{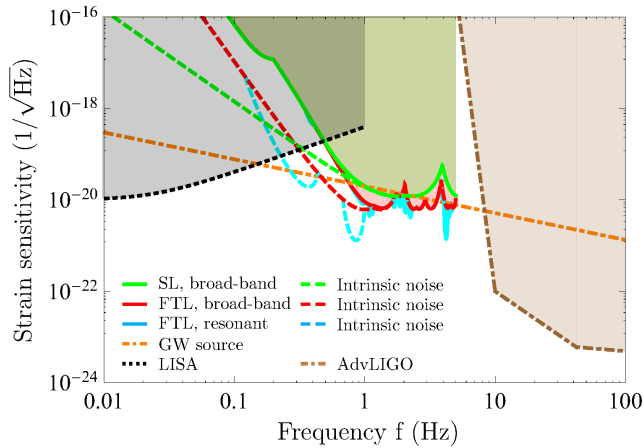}
\caption{(Color online) Spectral strain sensitivities of symmetric single (green lines) and multi-loop (red lines) interferometer compared to the strain induced by a black hole binary (orange dash-dotted line) taken from Ref.~\cite{Sesana2017JPCS}, including sky as well as polarization averaging~\cite{Moore2015CQG}.
Solid lines illustrate the effective detector sensitivity including vibrations of the retroreflection mirrors and Newtonian noise as discussed in Ref.~\cite{Chaibi2016PRD} assuming the isolation system from Ref.~\cite{Losurdo1999RSI}.
In the broad-band mode, both reach an intrinsic sensitivity (green dashed line, red dashed line) of better than $10^{-19}\,/\sqrt{\mathrm{Hz}}$ between 0.3\,Hz and 5\,Hz running three interferometer sequences interleaved with different free-evolution times $T$ (182\,ms, 234\,ms, 260\,ms).
The triple-loop can be extended to a resonant mode by increasing the number of atomic loops leading ideally to a higher intrinsic sensitivity at specific frequencies determined by $T$.
The nine-loop interferometer for $T=260\,\mathrm{ms}$ is about three times more sensitive at 0.85\,mHz (cyan dashed line).
The spectral response of our detectors fills the gap between the space-born detector LISA~\cite{AmaroSeoane2017arXiv} (black dotted line) and advanced LIGO~\cite{Abbott2016PRL_3,Moore2015CQG} (brown dash-dotted line).
}
\label{fig:strain_sensitivity_comp}
\end{figure}

Figure~\ref{fig:strain_sensitivity_comp} compares the spectral strain sensitivities, obtained by the broad-band mode of the single- (green lines) and triple-loop detector (red lines), and the narrow-band mode (cyan lines) of the multi-loop geometry, with a signal (orange dash-dotted line) generated by a black hole binary as discussed in Ref.~\cite{Sesana2017JPCS}.
They are also confronted with the anticipated strain sensitivity of the proposed space-borne detector LISA~\cite{AmaroSeoane2017arXiv} (black dotted line) and the operating advanced LIGO~\cite{Abbott2016PRL_3,Moore2015CQG} (brown dash-dotted line).
We emphasize that the sensitivity curves of our detector concepts fill the gap between LISA and advanced LIGO, enabling the terrestrial detection of infrasound gravitational waves.

With respect to the strain sensitivity, our designs share assumptions with other proposals~\cite{Canuel2017arXiv,Chaibi2016PRD} based on atom interferometry.
For two 10\,km long arms, we foresee an intrinsic total phase noise of $1\,${\textmu}$\mathrm{rad}/\sqrt{\mathrm{Hz}}$ achieved by $10^9$\,atoms starting every 100\,ms with an upward velocity of $gT$/2 where $g$ is the gravitational acceleration.
Moreover, we expect a 20\,dB sub-shot-noise detection~\cite{Hosten2016Nature}, a maximum of $T\sim260\,\mathrm{ms}$, and the symmetric transfer of 1000 photon recoils at the rubidium D2 line in each direction~\footnote{See Supplemental Material at [URL will be inserted by publisher] for parameter estimates.}, which is by a factor of 5 beyond current state of the art~\cite{Gebbe2019arXiv,Pagel2019arXiv,Jaffe2018PRL,Plotkin2018PRL,Kovachy2015Nature,McDonald2014EPL,Chiow2011PRL,Muller2009PRL,Clade2009PRL}.

To enable such a high efficiency for beam splitting and launching~\cite{Gebbe2019arXiv,Abend2016PRL,Szigeti2012NJP}, we require a residual atomic expansion rate of 100\,{\textmu}m/s corresponding to $\sim$100\,pK for $^{87}$Rb as achieved with delta-kick collimated ensembles of rubidium~\cite{Kovachy2015PRL} and Bose-Einstein condensates~\cite{Rudolph2016Diss,Muntinga2013PRL}.
Using Bose-Einstein condensates is compatible with the future utilization of entangled atoms~\cite{Kruse2016PRL}.

In addition, the assumed production rate of atomic ensembles enables an interleaved~\cite{Savoie2018ScAdv,Dutta2016PRL} operation of several interferometers.
Indeed, for the broad-band mode, we employ for the single- and triple-loop detector three interleaved interferometers with the free-fall times $T_1=T$, $T_2=0.9T$, and $T_3=0.7T$ to avoid peaks in the transfer function~\cite{Hogan2016PRA} resulting in the depicted intrinsic strain sensitivities (green, red dashed line).

Moreover, we extend the triple-loop scheme to a narrow-band mode~\cite{Graham2016PRD} featuring $3n$ loops, with a sensitivity enhanced by a factor $n\in\mathbb{N}$ at a specific frequency determined by the free-fall time $T$.
The cyan line in Fig.~\ref{fig:strain_sensitivity_comp} shows the sensitivity for the frequency 0.85\,mHz, corresponding to $T=260\,\mathrm{ms}$, and $n=3$.
The interleaved operation also ensures that the strain sensitivity features a high frequency cut-off at 5\,Hz.
\begin{table}[tb]
 \begin{center}
 \caption{Requirements on key parameters of the atomic source and launch mechanisms for the single-loop and triple-loop interferometer.
 Here, we list only the dominant contributions originating from the terrestrial gravity gradient and the Sagnac effect.
The instabilities in pointing, position $\delta x$ and velocity $\delta v_{x}$, $\delta v_{y}$ refer to 1\,s operation time which corresponds to 10 interferometry cycles.
We assume a maximum phase noise of 1\,{\textmu}rad in 1\,s with $k=2000\cdot2\pi/(780\,\mathrm{nm})$, and $T=0.26\,\mathrm{s}$.
Typical values for gravity gradients, the earth rotation, and the gravitational acceleration are $\Gamma=1.5\cdot10^{-6}\,1/\mathrm{s}^2$, $\Omega=5.75\cdot10^{-5}\,\mathrm{rad/s}$, and $g=9.81\,\mathrm{m/s}^2$, respectively.
The vertical distance between source and beam splitting zone is 30\,cm.
The relative relaunch pointing refers to an instability in the angle between the relaunch at $t=(9/5)T$ and $t=(21/5)T$.
A jitter in the pointing leads to a coupling to $\Gamma$ and $\Omega$.
The latter dominates for our choice of parameters.
The phase noise is given per arm.
Assuming no correlation, which is a valid assumption for shot noise, the noise would increase by a factor $\sqrt{2}$ in the differential signal between both arms.
The values in the upper part of the table assume no timing error for the relaunch.
The constraint set by a timing error shifting the relaunches by 10\,ns is shown at the bottom.}
 \begin{tabular}{p{0.50\columnwidth}p{0.22\columnwidth}p{0.22\columnwidth}}\hline
   & triple-loop & single-loop\\ \hline
 $\delta x$ from $\Gamma\cdot\delta x\cdot kT^{2}$  & n/a & 0.4\,nm \\
 -pointing launch & n/a & 1.4\,nrad \\
 $\delta v_{x}$ from $\Gamma\cdot\delta v_{x}\cdot kT^{3}$  & n/a & 1.7\,nm/s \\
 -pointing launch & n/a & 0.7\,nrad \\
 $\delta v_{y}$ from Sagnac term & n/a & 5.6\,pm/s (few~fK)\\
 -pointing launch & n/a & 2.2\,prad \\
 pointing $k$ (g matched to $10^{-7}$) & 48\,prad & 100\,prad \\
 relative relaunch pointing & 1\,prad & n/a \\ \hline  
 absolute relaunch pointing & 1\,nrad & n/a \\ \hline      
 \end{tabular}
 \label{tab:parameter}
 \end{center}
 \end{table}

Next to the atomic shot noise, other technical and environmental noise sources deteriorate the strain sensitivity.
Their impact is evaluated by the transfer function~\cite{Cheinet2008IEEE} of the interferometers.
Figure~\ref{fig:strain_sensitivity_comp} shows, that the intrinsic strain sensitivity of the single-loop (SL, green dashed line)~\cite{Chaibi2016PRD} and folded triple-loop (FTL, red dashed line) interferometer decays with $f^{-2}$ and $f^{-4}$ at lower frequencies. 
Below 1\,Hz, vibrations of the retroreflecting mirrors and Newtonian noise as modelled in Refs.~\cite{Chaibi2016PRD,Chaibi2016PRD_suppl,Harms2013PRD,Saulson1984PRD} limit the strain sensitivity (green, red, cyan solid lines).
Mirror vibrations enter in the differential signal of the interferometers due to the finite speed $c$ of light causing a delay of $2L/c\approx70\,${\textmu}$\mathrm{s}$, and are modelled with a differential weighting function~\cite{Cheinet2008IEEE}.
A suspension system isolating the retroreflection mirror against seismic noise~\cite{Chaibi2016PRD} according to the specifications of Ref.~\cite{Losurdo1999RSI} reduces this contribution to a level similar to the Newtonian noise.

We emphasize that the spectral sensitivity, in particular the low-frequency cut-off, can be tuned by the free-fall time~$T$ of the atoms.
In the case of the triple-loop interferometer the maximum value of~$T$ is only determined by the free-fall height.
For example, in a $\sim$1\,m-high chamber~$T$ can be tuned up to 260\,ms.

In contrast, in the single-loop interferometer, the tuning capability is restricted by the separation of the two twin lattices as illustrated in Fig.~\ref{fig:GWD_3_pics}~(b).
With $T$\,=\,260\,ms and adding 40\,ms for beam splitting, a height difference of $45\,\mathrm{cm}$ is needed.
In our scenario, the triple-loop detector surpasses the signal-to-noise ratio of the single-loop by a factor of about 2 for the broad-band, and about 9 for the narrow-band mode.

The relaunches introduce additional noise terms to the folded triple-loop interferometer.
The absolute pointing of the relaunch vectors can be adjusted with the interferometer itself by tuning their timings and their pointing. 
Implementations at the most quiet sites, featuring a residual rotation noise of $2\cdot10^{-11}\,\mathrm{(rad/s)/\sqrt{\mathrm{Hz}}}$~\cite{Schreiber2011PRL} enforce either measures such as mechanical noise dampening by one order of magnitude, or atom interferometric measurements sharing a relaunch pulse for two interleaved cycles~\cite{Savoie2018ScAdv,Dutta2016PRL,Meunier2014PRA,Biedermann2013PRL} to comply with the requirement on pointing noise.
Having the second relaunch of the first cycle as the first of a subsequent one leads to a behaviour of $\sim1/t$ for an integration time $t$.
For a white pointing noise, assuming a normal distribution with $\sigma=2\cdot10^{-11}\,\mathrm{rad}$ at 1\,s, the limit set by the intrinsic noise would be reached after an integration time of 400\,s.

In conclusion, present designs of terrestrial detectors of infrasound gravitational waves based on atom interferometry face several severe challenges related to scalability and atomic expansion necessitating femtokelvins temperatures which cannot be reached with today's atomic source concepts.
Therefore, we propose atomic folded-loop interferometers for horizontal antennas which overcome these stringent requirements at the cost of a very stable relaunch of the atoms.
In addition, they merge the advantages of horizontal detectors and vertical setups: As horizontal antennas, they display a scalability of the arm length, and do not rely on deep boreholes while preserving the tunability of the spectral response of vertical detectors enabling broadband and resonant detection modes.
Our scheme opens a new pathway to reach strain sensitivities of the order of $7\cdot10^{-21}\,/\sqrt{\mathrm{Hz}}$ at 1\,Hz in terrestrial detectors.

\begin{acknowledgements}
This work is supported by ``Nieders\"achsisches Vorab" through the ``Quantum- and Nano- Metrology (QUANOMET)" initiative within the project QT3 , the CRC 1227 DQmat within the project B07, the CRC 1128 geo-Q  within the project A02, the QUEST-LFS, the Deutsche Forschungsgemeinschaft (DFG, German Research Foundation) under Germany's Excellence Strategy - EXC-2123-B2, and the German Space Agency (DLR) with funds provided by the Federal Ministry for Economic Affairs and Energy (BMWi) due to an enactment of the German Bundestag under Grants No. DLR~50WM1556, 50WM1641, 50WM1952, 50WM1956, and 50WP1700. 
D.~S. gratefully acknowledges funding by the Federal Ministry of Education and Research (BMBF) through the funding program Photonics Research Germany under contract number 13N14875.
W.P.S. is grateful to Texas A\&M University for a Faculty Fellowship at the Hagler Institute of Advanced Study at Texas A\&M University, and to Texas A\&M AgriLife Research for the support of his work.
The Research of IQ$^{\text{ST}}$ is financially supported by the Ministry of Science, Research and Arts Baden-W{\"u}rttemberg.
\end{acknowledgements}

\end{document}


\title{\large{Supplemental material for}\normalsize\\ Scalable, symmetric atom interferometer for infrasound gravitational wave detection}

\def\affiqo  {\affiliation{Leibniz Universit\"at Hannover, Institut f\"ur Quantenoptik, Welfengarten 1, 30167 Hannover, Germany }}
\def\affulm {\affiliation{Institut f\"ur Quantenphysik and Center for Integrated Quantum Science and Technology \textnormal{(IQ$^{\text{ST}}$)}, Universit\"at Ulm, Albert-Einstein-Allee 11, D-89081 Ulm, Germany}}

\author{C. Schubert}
\email[]{schubert@iqo.uni-hannover.de}
\author{D. Schlippert}
\author{S. Abend}
\affiqo
\author{E. Giese}
\affiliation{Department of Physics, University of Ottawa, 25 Templeton Street, Ottawa, Ontario, Canada K1N 6N5}
\affulm
\author{A.~Roura}
\affulm
\author{W.~P.~Schleich}
\affulm
\affiliation{Institute of Quantum Technologies, German Aerospace Center (DLR), D-89069 Ulm, Germany}
\affiliation{Institute for Quantum Science and Engineering \textnormal{(IQSE)}, Texas A\&M AgriLife Research and Hagler Institute for Advanced Study, Texas A\&M
University, College Station, Texas 77843-4242, USA}
\author{W. Ertmer}
\author{E. M. Rasel}
\affiqo

\maketitle

\section{Parameter estimate for twin-lattice and rubidium}
Initially, sequential double Bragg diffraction~\cite{Ahlers2016PRL,Giese2013PRA} pulses couple the zero-momentum state to the $\pm\,8\,\hbar\,k_\mathrm{s}$ momentum states where $\lambda$ denotes the wavelength of the atomic transition and $k_\mathrm{s}{\equiv}2\pi/\lambda$ is the single-photon wave number.
This momentum separation ensures an efficient transfer to higher momentum states with subsequent Bloch oscillations induced by counter-propagating lattices~\cite{Gebbe2019arXiv}.
We expect a transfer of $100\,\hbar\,k_\mathrm{s}$ differential momentum in a millisecond, leading to a total duration of 20\,ms for a composite $\pi$/2-pulse.
The two trajectories separate with a velocity of $\hbar\,k/m_{\mathrm{Rb}}=2000\,\hbar\,k_\mathrm{s}/m_{\mathrm{Rb}}$ and are later deflected towards each other in a similar way in about 40\,ms with $(5/4)\,\hbar\,k/m_{\mathrm{Rb}}=2500\,\hbar\,k_\mathrm{s}/m_{\mathrm{Rb}}$ where $m_{\mathrm{Rb}}$ denotes the atomic mass of rubidium.
Symmetric de-acceleration by Bloch lattices, sequential double Bragg $\pi$-pulses, and symmetric acceleration by Bloch lattices invert the momentum.

In case of $^{87}$Rb $\lambda=780\,\mathrm{nm}$, the maximum wave packet separation would reach $\sim4.4$\,m. 

The additional relaunch, required in the folded triple-loop interferometer, can be based on the same combination of Bloch lattices and double Bragg diffraction with a duration of 15\,ms for the upward deflection~\cite{Abend2016PRL}.

\section{Coupling of timing errors and coherent relaunch}
The differential measurement scheme of our detectors suppresses timing errors of the beam splitters to first order, because they are common to both atom interferometers contributing to the signal.
On the contrary, the relaunches affect both atom interferometers individually.
Relaunches that are not centered around the intersections of the trajectories induce a phase shift if in addition the projection of the relaunch vector onto the effective wave vector of the beam splitters is nonzero.

We estimate the impact on the detector by determining the mean trajectory of the atoms and subsequent calculation of the phase shift~\cite{Borde2004GRG}.
Here, we neglect rotations and gravity gradients which are discussed later on without timing errors.
Cutting the mean trajectory into sections, our set of equations is
\begin{eqnarray*}
\begin{cases}
y_1(t) = y_0 + {v_0}t \hspace{5mm} \mathrm{for}\hspace{3mm}t<t_1 \\
y_2(t) = y_0 + {v_0}t + \alpha_1(a_1/2)(t_2-t_1)^2 + \alpha_1a_1(t_2-t_1)(t-t_2) \hspace{5mm} \mathrm{for}\hspace{3mm}t_2<t<t_3  \\
y_3(t) = y_0 + {v_0}t + \alpha_1(a_1/2)(t_2-t_1)^2 + \alpha_1a_1(t_2-t_1)(t_3-t_2) + \alpha_2(a_2/2)(t_4-t_3)^2 + \alpha_2a_2(t_4-t_3)(t-t_4)\\
\hspace{150mm}\mathrm{for}\hspace{3mm}t_4<t\\
\end{cases}
\end{eqnarray*}
with initial position $y_0$, initial velocity $v_0$, tilting angle deviation $\alpha_i$ from 90\,$^{\circ}$ with respect to the beam splitters, and acceleration during the relaunches $a_i$ with $i=1,2$.
Herein, $t_1$ marks the beginning of the first relaunch, $t_2$ its end, $t_3$ the beginning of the second relaunch, and $t_4$ its end, reflecting a finite duration of the relaunches.
The phase shift is estimated by
\begin{equation*}
\phi = k\left[y_1(0) - \frac{9}{4}y_1(T) + \frac{5}{2}y_2(3T) - \frac{9}{4}y_3(5T) + y_3(6T)\right]
\end{equation*}
with the pulse separation time $T$ and the effective wave number $k$, corresponding to the differential momentum between the two trajectories transferred during beam splitting.
We insert $t_1=(9/5)T-(\tau_1/2)+\delta\tau_1$, $t_1=(9/5)T+(\tau_1/2)+\delta\tau_1$, $t_3=(21/5)T-(\tau_2/2)+\delta\tau_2$, $t_4=(21/5)T+(\tau_2/2)+\delta\tau_2$ reflecting our interferometer.
Here, $\tau_1=t_2-t_1$ and $\tau_2=t_4-t_3$ denote the durations of the respective relaunch pulses, displaced by $\delta\tau_1$ and $\delta\tau_2$ in time from the intersection of the trajectories.
Consequently, the phase shift simplifies to
\begin{equation*}
\phi = \frac{5}{2}k(\alpha_2a_2\tau_2\delta\tau_2-\alpha_1a_1\tau_1\delta\tau_1).
\end{equation*}
We use the definitions $\Sigma\tau=\delta\tau_2+\delta\tau_1$ and $\Delta\tau=\delta\tau_2-\delta\tau_1$, assume equal duration $\tau_1=\tau_2=\tau$ and acceleration $a_1=a_2=a$ of the relaunches, to find the expression
\begin{equation}\label{eq:timing_error}
\phi = \frac{5}{4}k{\cdot}a\cdot\tau\left(\Sigma\tau\frac{\alpha_2-\alpha_1}{2}+\Delta\tau\frac{\alpha_2+\alpha_1}{2} \right).
\end{equation}
According to the first term, a common timing error leads to a phase shift if a relative tilt between the relaunch vectors is present.
For our parameters $k=2000\cdot2\pi/(780\,\mathrm{nm})$, $a\cdot\tau=(5/2)gT$, $T=0.26\,\mathrm{s}$, and $g=9.81\,\mathrm{m/s}^2$, this restricts a jitter in the differential angle $\Delta\alpha=\alpha_2-\alpha_1$ to 1\,nrad in 1\,s for a common timing error $\Sigma\tau$ of 10\,ns to keep this contribution to the phase noise below 1\,{\textmu}rad in 1\,s.
The second term in Eq.~\eqref{eq:timing_error} appears in case of a differential timing error and a nonzero mean tilt.
A differential timing error $\Delta\tau$ of 10\,ns implies the restriction of a common tilt $\Sigma\alpha=\alpha_2+\alpha_1$ to 1\,nrad.
Requirements on timings are within the capability of current real-time controllers.
The common tilt can be adjusted with the interferometer itself by scanning the tilt angles for several differential timings $\Delta\tau$ while zeroing $\Sigma\tau$.

\section{Requirements in a single-loop geometry}
Gravity gradients, rotations and other quantities can induce phase shifts in the differential signal of the two atom interferometers if the starting conditions are not matched.
Phase terms estimated according to Refs.~\cite{Hogan2008arXiv,Bongs2006APB,Borde2004GRG} to leading order in the limit of infinitely short pulses, and the resulting requirements for a single-loop interferometer are summarized in Table~\ref{tab:single_loop}.

\begin{table*}[b]
 \begin{center}
 \caption{Phase shifts and requirements on key parameters of the atomic source and launch mechanism for the single-loop interferometer. 
 The parameters in our estimation are: maximum phase noise of $\sigma\phi=1$\,{\textmu}rad in 1\,s, mean initial position $\delta x=r_m(10)$, $\delta r$, mean initial velocities $\delta v_{x}=v_m(10)$, $\delta v_{y}=v_m(10)$, $N=10^9\,\mathrm{atoms}$ injected every 0.1\,s, $k=2000\cdot2\pi/(780\,\mathrm{nm})$, $T=0.26\,\mathrm{s}$, $\Gamma=1.5\cdot10^{-6}\,1/\mathrm{s}^2$, $\Omega=5.75\cdot10^{-5}\,\mathrm{rad/s}$, gravitational acceleration $g=9.81\,\mathrm{m/s}^2$, distance between source and beam splitting zone $l=30\,\mathrm{cm}$, upward launch velocity $gT$, $\mathbf{k}\cdot\mathbf{g}/(|\mathbf{k}||\mathbf{g}|)\approx\beta$, difference in gravitational acceleration at the two atom interferometers ${\delta}g=10^{-7}\,g$, angle between second and first beam splitter $\delta{\beta}_2$, angle between last and second beam splitter $\delta{\beta}_3$.
 Numerical values are given per second.
}
 \begin{tabular}{p{0.35\textwidth}p{0.3\textwidth}p{0.32\textwidth}}\hline
   \textbf{Parameter} & \textbf{Formula} & \textbf{Numerical value}\\ \hline
 Variation of mean position (y direction) & $\delta y=\sigma\phi/\left( k\Gamma T^2\cdot\sqrt{2} \right)$ & $\delta y=4.3\cdot10^{-10}\,\mathrm{m}$ \\ %
 - Variation of launch angle (y direction) & $\delta \alpha_y=\sigma\phi/\left( k\Gamma T^2\cdot\sqrt{2}\cdot l \right)$ & $\delta \alpha_y=1.4\cdot10^{-9}\,\mathrm{rad}$ \\ %
 - Initial radius (y direction) & $\sigma_r=\delta y\cdot\sqrt{10\cdot N}$ & $\sigma_r=4.3\cdot10^{-5}\,\mathrm{m}$ \\ %
Variation of mean velocity (y direction) & $\delta v_{y}=\sigma\phi/(k\Gamma T^3\cdot\sqrt{2})$ & $\delta v_{y}=1.7\cdot10^{-9}\,\mathrm{m/s}$ \\ %
 - Variation of launch angle (y direction) & $\delta \alpha_{v,y}=\sigma\phi/(k\Gamma T^3\cdot\sqrt{2}\cdot gT)$ & $\delta \alpha_{v,y}=6.5\cdot10^{-10}\,\mathrm{rad}$ \\ %
 - Expansion rate (y direction) & $\sigma_v=\delta v_{y}\cdot\sqrt{10\cdot N}$ & $\sigma_v=1.7\cdot10^{-4}\,\mathrm{m/s}$ \\ %
 Variation of mean velocity (x direction) & $\delta v_{x}=\sigma\phi/(2k\Omega T^2\cdot\sqrt{2})$ & $\delta v_{x}=5.6\cdot10^{-12}\,\mathrm{m/s}$ \\ %
 - Variation of launch angle (x direction) & $\delta \alpha_{v,x}=\sigma\phi/(2k\Omega T^2\cdot\sqrt{2}\cdot gT/2)$ & $\delta \alpha_{v,x}=2.2\cdot10^{-12}\,\mathrm{rad}$ \\ 
 - Expansion rate (x direction) & $\sigma_v=\delta v_{x}\cdot\sqrt{10\cdot N}$ & $\sigma_v=5.6\cdot10^{-7}\,\mathrm{m/s}$ \\ %
 Difference in $g$, beam splitter pointing & $\sigma\phi=k{\delta}gT^2\left( \beta+\delta\beta_2+\delta\beta_3 \right)$ & ${\delta}g=10^{-7}\,g$, $\beta\approx\delta\beta_2\approx\delta\beta_3\approx10^{-10}\,\mathrm{rad}$ \\
 Initial position, beam splitter pointing & $\sigma\phi=k{\delta}r\left( -\delta\beta_2+\delta\beta_3 \right)$ & ${\delta}r=3.1\cdot10^{-7}\,\mathrm{m}$ \\
 - Initial radius & $\sigma_{r}=\delta r\cdot\sqrt{10\cdot N}$ & ${\sigma}_r=9.8\cdot10^{-2}\,\mathrm{m}$ \\
 Initial velocity, beam splitter pointing & $\sigma\phi=2kT{\delta}v \delta\beta_3$ & ${\delta}r=8.4\cdot10^{-7}\,\mathrm{m/s}$ \\
 - Expansion rate & $\sigma_{v}=\delta v\cdot\sqrt{10\cdot N}$ & ${\sigma}_v=2.7\cdot10^{-1}\,\mathrm{m/s}$ \\ 
 \hline  
 \end{tabular}
 \label{tab:single_loop}
 \end{center}
 \end{table*}
 
Ideally, instabilities in the starting conditions are shot noise limited.
Assuming that other noise contributions can be neglected, the instability of the mean position $r_m(n)=\sigma_r/\sqrt{n{\cdot}N}$ and velocity $v_m(n)=\sigma_v/\sqrt{n{\cdot}N}$ of the wave packet for $n$ cycles depend on the number $N$ of atoms per cycle, the initial radius $\sigma_r$, and expansion rate $\sigma_v$ of the wave packet.

Gravity gradients $\Gamma$ impose a limit on the instabilities of the center and the velocity of the wave packet to $4.3\cdot10^{-10}\,\mathrm{m}$ at 1\,s and $1.7\cdot10^{-9}\,\mathrm{m/s}$ at 1\,s, respectively, pointing instabilities of the initial launch to $\sim10^{-9}\,\mathrm{rad}$ at 1\,s, constraining initial radius and expansion rate of the wave packet to $4.3\cdot10^{-5}\,\mathrm{m}$ and $1.7\cdot10^{-4}\,\mathrm{m/s}$ for $10^{10}$ atoms per second, respectively.
If the gravity gradients are equal and known at the two light-pulse atom interferometers interrogated by the same laser beam, an adjustment of the wave number at the second pulse can be introduced to relax these requirements~\cite{Roura2017PRL}.

Significantly more stringent conditions are set by the Sagnac effect caused by the rotation of the earth $\Omega$ which restricts instabilities in the pointing of the initial launch to the level of $2.2\cdot10^{-12}\,\mathrm{rad}$ at 1\,s, and in the center of wave packet to $5.6\cdot10^{-12}\,\mathrm{m/s}$ at 1\,s, implying a residual expansion rate of at most $5.6\cdot10^{-7}\,\mathrm{m/s}$ corresponding to few femtokelvins. 
The latter is challenging even for delta-kick collimated, dilute Bose-Einstein condensates.

Beam splitters which are not exactly perpendicular to gravity by an angle $\beta$ induce a phase shift in a single interferometer that is suppressed in the differential signal.
If the gravitational acceleration $g$ differs at the locations of the two atom interferometers by ${\delta}g$, a spurious phase shift remains.
Assuming ${\delta}g=10^{-7}\,g$, instabilities in $\beta$ have to be limited to $10^{-10}\,\mathrm{rad}$.
A misalignment at the level of $10^{-10}\,\mathrm{rad}$ implies less stringent requirements on initial radius and expansion of the wave packet than gravity gradients and rotations, even when assuming a compensation of gravity gradients~\cite{Roura2017PRL} by a factor of 1000.

In addition, instabilities in density shifts $\propto\delta\,N/(\sigma_\mathrm{r})^3$ for an ensemble with $N$ atoms and radius $\sigma_\mathrm{r}$ restrict the maximum initial density and fluctuations in the beam splitting fidelity $\delta$ which affects the relative density of atoms in the two arms~\cite{Debs2011PRA,Dimopoulos2008PRD}.
Assuming $\delta\sim3\cdot10^{-5}$, a compensation of the phase shift induced by gravity gradients~\cite{Roura2017PRL} has to be implemented to at least a factor of 100 to simultaneously comply with the position requirement scaling as $\propto\sigma_\mathrm{r}/\sqrt{N}$.

\section{Requirements in a triple-loop geometry}
\begin{table*}[t]
 \begin{center}
 \caption{Phase shifts and requirements on key parameters of the atomic source and launch mechanisms for the triple-loop interferometer.
 The angle between the two subsequent relaunches is given by $\Delta\alpha$, the upward launch velocity by $gT/2$, variations $\delta\beta_i$ in the angle between the current beam splitter $i$ and the previous one.
 Other parameters are the same as in Table~\ref{tab:single_loop}.
 Requirements on the mean position ${\delta}y$ are less demanding than on the mean position ${\delta}x$.
}
 \begin{tabular}{p{0.35\textwidth}p{0.36\textwidth}p{0.26\textwidth}}\hline
   \textbf{Parameter} & \textbf{Formula} & \textbf{Numerical value}\\ \hline
 Relaunch pointing (y direction) & $\Delta\alpha=\sigma\phi/\left[-(39/10) k \Gamma g T^4 \cdot\sqrt{2}\right]$ & $\Delta\alpha=3.3\cdot10^{-10}\,\mathrm{rad}$ \\ %
 Relaunch pointing (x direction) & $\Delta\alpha=\sigma\phi/\left( 9 k g T^3 \Omega \cdot\sqrt{2}\right)$ & $\Delta\alpha=10^{-12}\,\mathrm{rad}$ \\ %
 Variation of mean position (y direction) & ${\delta}y=\sigma\phi/\left[ (15/4) \Gamma ^2 k T^4 \cdot\sqrt{2}\right]$ & ${\delta}y=1.1\cdot10^{-3}\,\mathrm{m}$ \\ %
 - Variation of launch angle (y direction) & $\delta \alpha_y=\sigma\phi/\left[ (15/4) \Gamma ^2 k T^4\cdot\sqrt{2}\cdot l \right]$ & $\delta \alpha_y=3.8\cdot10^{-3}\,\mathrm{rad}$ \\ %
 - Initial radius (y direction) & $\sigma_{r}=\delta y\cdot\sqrt{10\cdot N}$ & $\sigma_{r}=1.1\cdot10^{2}\,\mathrm{m}$ \\ %
 Variation of mean position (y direction) & ${\delta}y=\sigma\phi/\left[ (45/4) k T^4 \Omega^4 \cdot\sqrt{2}\right]$ & ${\delta}y=7.8\cdot10^{1}\,\mathrm{m}$ \\ %
 - Variation of launch angle (y direction) & $\delta \alpha_y=\sigma\phi/\left[ (45/4) k T^4 \Omega^4\cdot\sqrt{2}\cdot l \right]$ & $\delta \alpha_y=2.6\cdot10^{2}\,\mathrm{rad}$ \\ %
 - Initial radius (y direction) & $\sigma_r=\delta y\cdot\sqrt{10\cdot N}$ & $\sigma_r=7.8\cdot10^{6}\,\mathrm{m}$ \\ %
 Variation of mean velocity (y direction) & ${\delta}v_y=\sigma\phi/\left[ (45/4) \Gamma ^2 k T^5 \cdot\sqrt{2}\right]$ & ${\delta}v_y=1.5\cdot10^{-3}\,\mathrm{m/s}$ \\ %
 - Variation of launch angle (y direction) & $\delta \alpha_{v,y}=\sigma\phi/(\cdot\sqrt{2}\cdot gT/2)$ & $\delta \alpha_{v,y}=1.1\cdot10^{-3}\,\mathrm{rad}$ \\ %
 - Expansion rate (y direction) & $\sigma_v=\delta v_{y}\cdot\sqrt{10\cdot N}$ & $\sigma_v=1.5\cdot10^{2}\,\mathrm{m/s}$ \\ %
 Variation of mean velocity (x direction) & ${\delta}v_x=\sigma\phi/\left( 15 k T^4 \Omega ^3 \cdot\sqrt{2}\right)$ & ${\delta}v_x=3.4\cdot10^{-3}\,\mathrm{m/s}$ \\ %
 - Variation of launch angle (x direction) & $\delta \alpha_{v,x}=\sigma\phi/(15 k T^4 \Omega ^3\cdot\sqrt{2}\cdot gT/2)$ & $\delta \alpha_{v,x}=2.6\cdot10^{-3}\,\mathrm{rad}$ \\ %
 - Expansion rate (x direction) & $\sigma_v=\delta v_{x}\cdot\sqrt{10\cdot N}$ & $\sigma_v=3.4\cdot10^{2}\,\mathrm{m/s}$ \\ %
 Variation of mean velocity (x direction) & ${\delta}v_x=\sigma\phi/\left[ (15/4) \Gamma k T^4 \Omega \cdot\sqrt{2}\right]$ & ${\delta}v_x=3\cdot10^{-5}\,\mathrm{m/s}$ \\ %
 - Variation of launch angle (x direction) & $\delta \alpha_{v,x}=\sigma\phi/\left[ (15/4) \Gamma k T^4 \Omega\cdot\sqrt{2}\cdot gT/2 \right]$ & $\delta \alpha_{v,x}=2.3\cdot10^{-5}\,\mathrm{rad}$ \\ %
 - Expansion rate (x direction) & $\sigma_v=\delta v_{x}\cdot\sqrt{10\cdot N}$ & $\sigma_v=3\,\mathrm{m/s}$ \\ %
 Difference in $g$, beam splitter pointing & $\sigma\phi=(9/8)k{\delta}gT^2\left( \delta\beta_3-9\delta\beta_4+16\delta\beta_5 \right)$ & ${\delta}g=10^{-7}\,g$, $\delta\beta_3/16=\delta\beta_4/9=\delta\beta_5=4.8\cdot10^{-11}\,\mathrm{rad}$ \\
 Initial position, beam splitter pointing & $\sigma\phi=(1/4)k{\delta}r\left( -4\delta\beta_2+5\delta\beta_3-5\delta\beta_4+4\delta\beta_5 \right)$ & ${\delta}r=1.8\cdot10^{-7}\,\mathrm{m}$, $(5/4)\delta\beta_2=(5/4)\delta\beta_5=\delta\beta_3=\delta\beta_4=10^{-10}\,\mathrm{rad}$\\
 - Initial radius & $\sigma_{r}=\delta r\cdot\sqrt{10\cdot N}$ & $\sigma_{r}=1.8\cdot10^{-2}\,\mathrm{m}$ \\
 Initial velocity, beam splitter pointing & $\sigma\phi=(3/4)kT{\delta}v\left( 3\delta\beta_3-7\delta\beta_4-8\delta\beta_5 \right)$ & ${\delta}v=1.6\cdot10^{-7}\,\mathrm{m/s}$, $(3/8)\delta\beta_3=(7/8)\delta\beta_4=\delta\beta_5=10^{-10}\,\mathrm{rad}$\\
 - Expansion rate & $\sigma_{v}=\delta v\cdot\sqrt{10\cdot N}$ & $\sigma_{v}=1.6\cdot10^{-2}\,\mathrm{m}$ \\
 \hline  
 \end{tabular}
 \label{tab:triple_loop}
 \end{center}
 \end{table*}
Analogous to the previous section, we now estimate phase terms according to Refs.~\cite{Hogan2011GRG,Hogan2008arXiv,Bongs2006APB,Borde2004GRG} to leading order for a triple-loop interferometer, deduce the requirements, and summarize them in Table~\ref{tab:triple_loop}.

Compared to a single-loop arrangement, a coupling of the starting conditions to rotations and accelerations only appears in higher order terms, and leads to requirements which are relaxed by several orders of magnitude.
Beam splitter tilts imply tighter bounds on the initial radius and expansion of the wave packet of $1.8\cdot10^{-2}\,\mathrm{m}$ and $1.6\cdot10^{-2}\,\mathrm{m/s}$, still less stringent by orders of magnitude than for the single-loop geometry.
Due to the modest requirement on the initial radius, the initial density can be adjusted to reduce the impact of interactions~\cite{Debs2011PRA,Dimopoulos2008PRD}.
High beam splitting efficiency~\cite{Gebbe2019arXiv,Abend2016PRL,Szigeti2012NJP} will set a tighter requirement on the expansion rate, estimated to be $10^{-4}$\,m/s.

Pointing jitter of the relaunches of the folded triple-loop geometry breaks the symmetry and leads to spurious phase shifts, depending on gravity gradients and rotations.
For our parameter set, rotations impose a limit on the jitter between the two subsequent relaunches of $10^{-12}\,\mathrm{rad}$ at 1\,s.